\documentclass{ws-procs9x6}

\def\ga{\gamma}

\def\Ga{\Gamma}

\def\mn{{\mu\nu}}

\def\frac#1#2{{\textstyle{{#1}\over {#2}}}}

\def\lsim{\mathrel{\rlap{\lower4pt\hbox{\hskip1pt$\sim$}}
    \raise1pt\hbox{$<$}}}
\def\gsim{\mathrel{\rlap{\lower4pt\hbox{\hskip1pt$\sim$}}
    \raise1pt\hbox{$>$}}}
\def\sqr#1#2{{\vcenter{\vbox{\hrule height.#2pt
         \hbox{\vrule width.#2pt height#1pt \kern#1pt
         \vrule width.#2pt}
         \hrule height.#2pt}}}}

\newcommand{\beq}{\begin{equation}}
\newcommand{\eeq}{\end{equation}}
\newcommand{\bea}{\begin{eqnarray}}
\newcommand{\eea}{\end{eqnarray}}

\begin{document}

\title{QUATERNIONIC FORMULATION OF THE\\
 DIRAC EQUATION}

\author{D.\ COLLADAY,$^*$ P.\ MCDONALD, and D.\ MULLINS}

\address{New College of Florida\\
Sarasota, FL 34243, USA\\
$^*$E-mail: colladay@ncf.edu}

\begin{abstract}
The Dirac equation with Lorentz violation involves additional coefficients 
and yields a fourth-order polynomial that must be solved to yield the dispersion
relation.
The conventional method of taking the determinant of $4\times 4$ matrices of 
complex numbers often yields unwieldy dispersion relations.
By using quaternions, the Dirac equation may be reduced to $2 \times 2$ form
in which the structure of the dispersion relations become more transparent.
In particular, it is found that there are two subsets of Lorentz-violating parameter
sets for which the dispersion relation is easily solvable.  Each subset contains half of the
parameter space so that all parameters are included.

\end{abstract}

\bodymatter
\section{General introduction}
The concept of searching for small remnant Lorentz- and CPT-violation was initially 
motivated in low energy limits of string theory, but has since branched out to include
much more general underlying theories of Nature \cite{kps}.
Specific coefficients to parameterize the theory were proposed 
and were later generalized slightly to allow for terms violating the gauge invariance of
the Standard Model \cite{ck}.
The resulting Dirac equation is modified due to the presence of the constant background 
vector and tensor fields.
The first general expression for the covariant dispersion relation involving all of the coefficients 
at the same time was initially found using the determinant of a $4\times4$ matrix operator
\cite{ralph1}.
The full expression is rather unwieldy as it leads to a fourth-order polynomial in the 
energy and momentum variables.
In this work, we use quaternions to reduce the full $4 \times 4$ matrix Dirac operator
to $2 \times 2$ form.
The mathematical properties of the quaternions are then used to obtain the dispersion
relation in a more tractable form \cite{cmm}.

\section{Some introduction to quaternions}
The quaternions are generated by four basis elements denoted $1, \hat i, \hat j, \hat k$,
with multiplicative properties
 \beq
  \hat i ^2 = \hat j^2 = \hat k^2 = -1,
  \eeq
 \beq
 \hat i \cdot \hat j =  - \hat j \cdot \hat i = \hat k, 
 \eeq
 and cyclic permutations.
 The quaternions are particularly interesting since they satisfy the axioms 
 for a mathematical skew field.
 A field is an algebraic structure with notions of addition, subtraction, multiplication,
 and division satisfying various properties, the key of which is that there are no nontrivial
 zero divisors.  
 The real numbers provide the simplest example with one dimension, they are commutative, and 
 have a natural ordering.
 The complex numbers are a two-dimensional example, but they lose the natural ordering
property that the real numbers have.
 The quaternions are a four-dimensional field, but they are noncommutative.
 Octonions are an eight-dimensional generalization, but there associativity is also lost
 making them particularly cumbersome to deal with.
 Frobenius presented a famous theorem in 1877 that proves the real, complex, and quaternionic
 numbers are in fact the only finite dimensional, associative division algebras.
 
 The quaternions may be represented using the Pauli sigma matrices as
 \beq
i\sigma_1  \longrightarrow  \hat{k}, \hspace{.25in} i\sigma_2
\longrightarrow  \hat{j},  \hspace{.25in} i\sigma_3 \longrightarrow
\hat{i}.
\eeq
This furnishes an explicit $2 \times 2$ representation for the quaternions as
\beq
 q = q_0 + q_1 \hat i + q_2 \hat j + q_3 \hat k = \left( 
\begin{matrix}
  q_0 + i q_1  & q_2 + i q_3 \cr
  -q_2 + i q_3 & q_0 - i q_1 
 \end{matrix} 
\right) .
\eeq
The name pure imaginary quaternion is given to a quaternion of the form
$\hat q = q_1 \hat i + q_2 \hat j + q_3 \hat k$ where the pure real piece vanishes.
Pure imaginary quaternions satisfy the following useful relations:
\begin{itemize}
\item $\hat a \hat b + \hat b \hat a = - 2 \vec a \cdot \vec b $,
\item $\hat a \hat b - \hat b \hat a = -2 \hat{(  a \times  b)}$,
\item $\hat a \hat b \hat c - \hat c \hat b \hat a = 2 \vec a \cdot (\vec b \times \vec c)$,
\end{itemize}
where $\hat {(a \times b)}$ indicates the quaternion that results after applying the
conventional cross product.

Another useful interpretation of the unit quaternions involves their action as generators
of rotations.
If $q = \cos({\theta/2}) - \hat q \sin({\theta /2 })$, then $\hat q \hat a \hat q^{-1} = \hat a^\prime$
where $\hat a^\prime = \hat a_\perp \cos{\theta} + \hat{( q \times a)} \sin{\theta} + \hat a_{||}$
is the quaternion corresponding to the vector $\hat a$ rotated by an angle $\theta$ about $\hat q$.
The perpendicular and parallel components of $\hat a$ are defined with respect to $\hat q$.
Such an interpretation makes quaternions useful in areas such as 3D gaming and spacecraft attitude controls.

\section{Quaternionic form of the Dirac equation}
The gamma matrices in the Dirac representation take the convenient block form
\beq
\vec \gamma = \left( 
\begin{matrix}
0 & \vec \sigma \cr
- \vec \sigma & 0
\end{matrix}
\right), \gamma^0 = \left( 
\begin{matrix}
1 & 0 \cr
0 & -1
\end{matrix}
\right).
\eeq
Using quaternions, the conventional free Dirac equation can be expressed as
\beq\left[ \left( \begin{matrix}
  p_0-m  & 0 \cr
  0 & -p_0-m 
\end{matrix} \right) +i \left( \begin{matrix}
  0 & \hat{p} \cr
  -\hat{p} & 0 
\end{matrix}\right) \right] \left(\begin{matrix}
  \phi \cr
  \xi 
\end{matrix}\right) = \left( \begin{matrix}
 0 \cr
0
  \end{matrix} \right) ,
\eeq
where
\beq
\hat{p} = \sum_j ip^j\sigma^j = p_3\hat{i} +p_2\hat{j} + p_1\hat{k}
\eeq
is a pure imaginary quaternion.
The lower row gives the spinor solution as
\beq
\xi = - {i \hat p \over p^0 + m} \phi,
\eeq
and the upper row then reproduces the dispersion relation
\beq
p_0^2 - \vec p~^2  = m^2.
\eeq
Note that $\phi$ and $\chi$ are two-component complex spinors 
that serve as a module on which the 
quaternions act.
This is in contrast to other approaches to using quaternionic valued wave functions
as has been attempted several times in the literature. \cite{quatdirac}.

\section{Quaternionic form of the perturbed Dirac equation}
The Dirac equation with Lorentz violation takes the form
\beq
(\Gamma^\mu p_\mu - M) \psi = 0,
\eeq
where
\beq
\Gamma^\nu  =  \gamma^\nu + c^{\mu\nu}\gamma_\mu + d^{\mu\nu}\gamma_5 \gamma_\mu
+ e^\nu + if^{\nu}\gamma_5 + \frac12 g^{\lambda \mu \nu}
\sigma_{\lambda \mu} ,
\eeq
and
\beq
M  =  m + a_\mu\gamma^\mu + b_\mu\gamma_5 \gamma^\mu 
+ \frac12 H_{\mu\nu}\sigma^{\mu\nu}.
\eeq
The coefficients $a_\mu$, $b_\mu$, etc... are Lorentz-violating constant background
fields.

Some field redefinitions can be used to simplify the model\cite{cmcd}. The first \cite{lehnert} is to 
fix $\Ga^0 \rightarrow \ga^0$ using the transformation $\psi = A \chi$, 
$A = (\ga^0\Ga^0)^{-1/2}$.  This makes the hamiltonian hermitian and is often a necessary
first step in properly interpreting the Lorentz-violating physical effects of any given
experiment.
The second involves a linear transformation on the momentum and mass space
to remove $a^\mu$, $c^\mn$, and $e^\mu$ terms
\bea
\hat p^\prime & = & \hat p - \hat a - \hat c_p , \\
p_0^\prime &=& (1 + c^{00}) p_0 - a_0 - \vec c_1 \cdot \vec p , \\ 
m^\prime& =& m - \vec e \cdot \vec p . \quad 
\eea
Since the transformation is linear, it is simple to invert at the end of the calculation.
To avoid cumbersome notation, the primes are dropped in the rest of the calculations.
Finally, the $f^\mu$ term can be removed using a more complicated redefinition \cite{altschul}.

The perturbed Dirac equation can then be put into quaternionic form
\beq
\left[ \left(\begin{matrix}
  p_0 - m  & \alpha_0+\hat{\alpha} \cr
  -\alpha_0 +\hat{\alpha} & -p_0 - m 
\end{matrix}\right)  +i \left(\begin{matrix}
  \hat{\epsilon} & -\hat{p} \cr
  \hat{p} & \hat{\delta} 
\end{matrix}\right) \right] \left(\begin{matrix}
  \phi \cr
  \xi 
\end{matrix}\right) = \left( \begin{matrix} 
0 \cr
0  
\end{matrix} \right).
 \label{dirac2.2}
\eeq
The parameters in the Dirac operator with hats are pure imaginary quaternions,
while the unhatted objects are real.
Their relation to the original Lorentz-violating coefficients are given by
\begin{eqnarray}
\alpha_0  =  b^0 + \hat{d}_1\cdot \hat{p},   & \hspace{.25in}& \hat{\alpha}  = 
\hat{H} - \hat{G}, \label{alphadef3.1} \\
\hat{\epsilon}  =  \hat{b} +\hat{d}_p +(\hat{g}-\hat{h}), & \hspace{.25in} & 
\hat{\delta}  =  -\hat{b} -\hat{d}_p +
(\hat{g}-\hat{h}), 
\label{epsilondef3.1}  
\end{eqnarray}
with the notation
\begin{eqnarray*}
\hat{d}^i_1   =  d^{0i},  & \hspace{.25in}& 
 \hat{d}^i_p  =  d^{ij}p_j,  \\ 
\hat{H}^i   =  H^{0i},  & \hspace{.25in}& 
\hat{G}^i  =  g^{0ij}p_j,  \\ 
\hat{h}^i   =  1/2 \epsilon^{ijk}H^{jk},  & \hspace{.25in}& 
\hat{g}^i  =  1/2 \epsilon^{ijk}g^{jkl}p_l.
\end{eqnarray*}
Substituting the first row equation of Eq.\ \eqref{dirac2.2} into the second row yields
\beq
\left[(\alpha_0 - \hat{\alpha} +
i\hat{p}) (x-i\hat{\epsilon})(\alpha_0 + \hat{\alpha} +
i\hat{p}) + r(-y+i\hat{\delta})\right] \xi = 0 ,
\eeq
where $x = p^0 - m$, $y = p^0 + m$, and $r= x^2 -|\hat{\epsilon}|^2$.
This equation reduces to the form
\beq
(q_1 +i\hat q_2)\xi = 0,
\eeq
where $q_1$ is real and $\hat q_2$ is a pure-imaginary quaternion.
This equation yields the dispersion relation from the eigenvalue condition
 $|\hat q_2^2| = q_1^2$.
A fourth-order polynomial in the energy $p^0$ results in the form
\beq
\beta_0 + \beta_1 p_0 + \beta_2 (p_0)^2 + (p_0)^4 = 0.
\eeq
The linear term $\beta_1$ vanishes for special parameter choices leading to a simple 
factorization of the dispersion relation.

Case 1 involves setting $\hat \epsilon = \hat \delta$, $\hat \alpha =0$.  This is equivalent
to setting $\hat b = \hat d_p = \hat H = \hat G = 0 $, while leaving the other parameters 
arbitrary.
The resulting dispersion relation takes the form
\beq
p_0^2 =   \vec p^2 + m^2 + \alpha_0^2 + \vec \delta^2 \pm 2 \sqrt{D_1(\vec p)} ,
\eeq
where 
\beq
D_1(\vec p) = (\vec \delta \times \vec p)^2 + (\alpha_0 \vec p  -  m \vec \delta)^2.
\eeq

Case 2 involves setting $\hat \epsilon = - \hat \delta$, $\alpha_0 = 0$.  This is equivalent
to setting $\hat h = \hat g = b_0 = \hat d_1 = 0$, while leaving the other parameters arbitrary.
The resulting dispersion relation takes the form
\beq
p_0^2 =   \vec p^2 + m^2 + \vec \alpha^2+ \vec \delta^2 \pm 2 \sqrt{D_2(\vec p)} ,
\eeq
where
\beq
D_2(\vec p)  =  (\vec \alpha \times \vec p - m \vec \delta)^2 
 + (\vec \delta \cdot \vec p)^2 + (\vec \alpha \cdot \vec \delta)^2.
\eeq


\begin{thebibliography}{xx}

\bibitem{kps}
V.A.\ Kosteleck\'y and S.\ Samuel,
Phys.\ Rev.\ D {\bf 39}, 683 (1989);
Phys.\ Rev.\ D {\bf 40}, 1886 (1989);
Phys.\ Rev.\ Lett.\ {\bf 63}, 224 (1989);
Phys.\ Rev.\ Lett.\ {\bf 66}, 1811 (1991);
V.A.\ Kosteleck\'y and R.\ Potting,
Nucl.\ Phys.\ B {\bf 359}, 545 (1991);
Phys.\ Lett.\ B {\bf 381}, 89 (1996);
Phys.\ Rev.\ D {\bf 63}, 046007 (2001); 
V.A.\ Kosteleck\'y, M.\ Perry, and R.\ Potting,
Phys.\ Rev.\ Lett.\ {\bf 84}, 4541 (2000). 

\bibitem{ck} 
D.\ Colladay and V.A.\ Kosteleck\'y,
Phys.\ Rev.\ D {\bf 55}, 6760 (1997);
Phys.\ Rev.\ D {\bf 58}, 116002 (1998).

\bibitem{ralph1}
V.A.\ Kosteleck\'y and R.\ Lehnert,
Phys.\ Rev.\ D {\bf 63}, 065008 (2003).

\bibitem{cmm}
D.\ Colladay, D.\ Mullins, and P.\ McDonald,
J.\ Phys.\ A {\bf 43}, 275202 (2010).

\bibitem{quatdirac}
See, for example,
D.\ Schuricht and M.\ Greiter,
Eur.\ J.\ Phys.\ {\bf 25} 755 (2004).

\bibitem{cmcd}
D.\ Colladay and P.\ McDonald,
J.\ Math.\ Phys.\ {\bf 43} 3554 (2002).

\bibitem{lehnert}
R.\ Lehnert,
J.\ Math.\ Phys.\ {\bf 45} 3399 (2004).

\bibitem{altschul}
B.\ Altschul,
J.\ Phys.\ A {\bf 39},13757 (2006).


\end{thebibliography}
\end{document}